\documentclass[11pt]{article}
\usepackage{moriond}

\usepackage{amsmath}
\usepackage{graphicx}

\newcommand{\Photo}{\includegraphics[height=35mm]{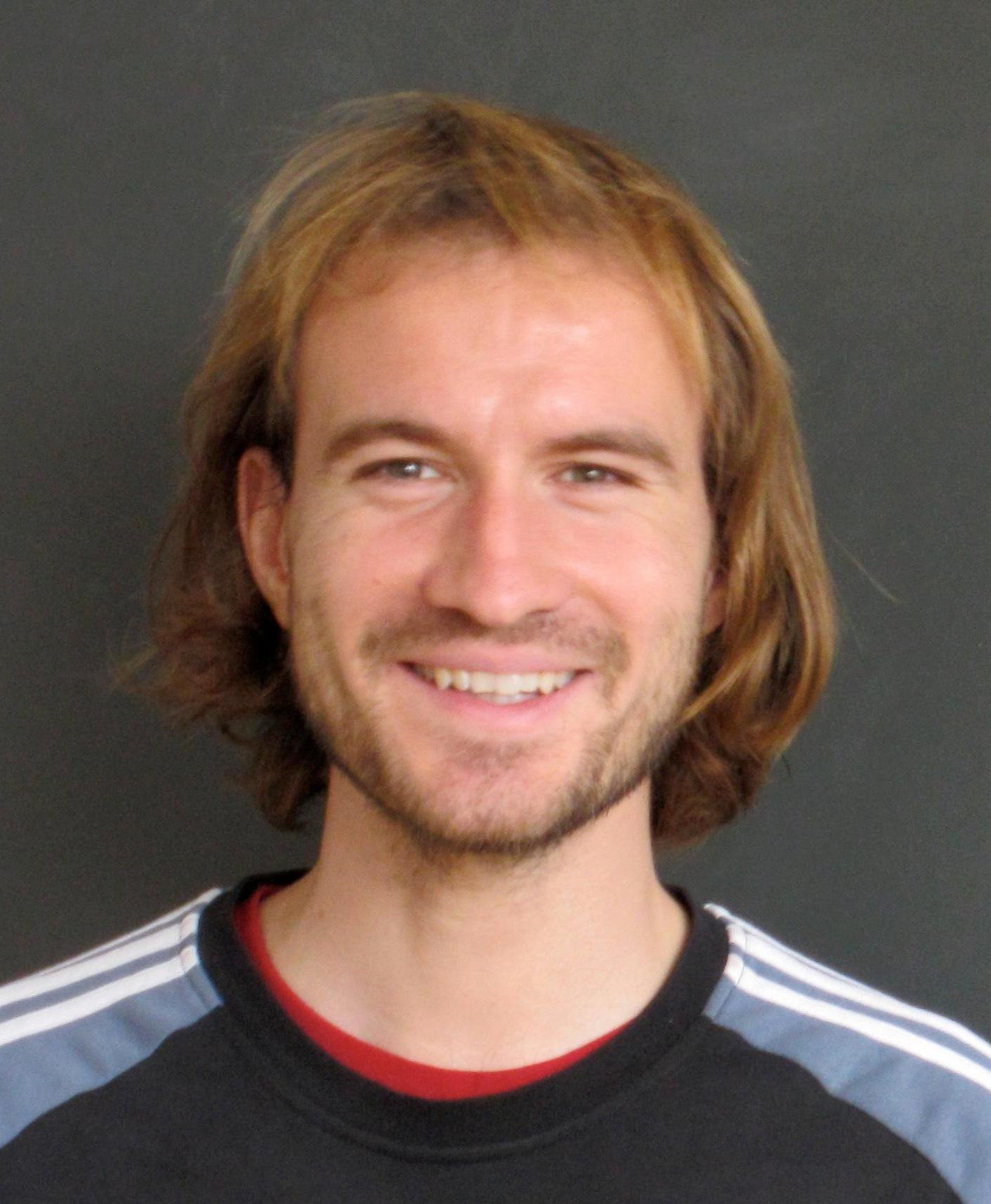}}

\begin{document}
%\textbox{PSI-PR-16-05}
\vspace*{4cm}
\title{New Physics in the Flavour Sector}

\author{ANDREAS CRIVELLIN}
\address{Paul Scherrer Institut, CH--5232 Villigen PSI, Switzerland}

\maketitle\abstracts{Several experiments observed deviations from the Standard Model (SM) in the flavour sector: LHCb found a $4-5\,\sigma$ discrepancy compared to the SM in $b\to s\mu^+\mu^-$ transitions (recently supported by an Belle analysis) and CMS reported a non-zero measurement of $h\to\mu\tau$ with a significance of $2.4\,\sigma$. Furthermore, BELLE, BABAR and LHCb founds hints for the violation of flavour universality in $B\to D^{(*)}\tau\nu$. In addition, there is the long-standing discrepancy in the anomalous magnetic moment of the muon. Interestingly, all these anomalies are related to muons and taus, while the corresponding electron channels seem to be SM like. This suggests that these deviations from the SM might be correlated and we briefly review some selected models providing simultaneous explanations.}

%%%%%%%%%%%
\section{Introduction}
\label{intro}
%%%%%%%%%%%

The discovery the Higgs at the LHC provided the final ingredient of the SM. While no direct evidence for physics beyond the SM was found during the first LHC run, there are some interesting indirect hints for NP in the flavor sector, mainly in semileptonic decays of $B$-mesons, the SM-forbidden decay $h\to\mu\tau$ of the Higgs boson and the long-lasting discrepancy in the anomalous magnetic moment (AMM) of the muon\footnote{We do not discuss the anomaly in $\epsilon^\prime/\epsilon$~\cite{Buras:2015yba} here for which possible solutions include $Z'$ bosons~\cite{Buras:2015kwd} or the MSSM~\cite{Kitahara:2016otd}.}.

{\boldmath $b\to s\ell^+\ell^-$:} Deviations from the SM found by LHCb~\cite{LHCb:2015dla} in the decay $B\to K^* \mu^+\mu^-$ arise mainly in an angular observable called $P_5^\prime$~\cite{Descotes-Genon:2013vna}, with a significance of $2$--$3\sigma$ depending on assumptions made for the hadronic uncertainties~\cite{Descotes-Genon:2014uoa,Altmannshofer:2014rta,Jager:2014rwa}. This measurement recently received support from a (less precise) BELLE measurement~\cite{Abdesselam:2016llu}. In the decay $B_s\to\phi\mu^+\mu^-$, LHCb also uncovered~\cite{Aaij:2015esa} deviations compared to the SM prediction from lattice QCD~\cite{Horgan:2013pva,Horgan:2015vla} of $3.5\sigma$ significance~\cite{Altmannshofer:2014rta}. LHCb has further observed lepton flavor universality violation (LFUV) in $B\to K\ell^+\ell^-$ decays~\cite{Aaij:2014ora} across the dilepton invariant-mass-squared range $1\,{\rm GeV}^2<m_{\ell\ell}^2<6\,{\rm GeV}^2$.  Here, the measured ratio branching fraction ratio 
$
R(K)=\frac{{\rm Br}[B\to K \mu^+\mu^-]}{{\rm Br}[B\to K e^+e^-]}
$
disagrees with the theoretically clean SM prediction by $2.6\sigma$. Combining these observables with other $b\to s$ transitions, it is found that NP is preferred over the SM by $4$--$5\sigma$~\cite{Altmannshofer:2015sma,Descotes-Genon:2015uva,Hurth:2016fbr}. 

{\boldmath $B\to D^{(*)}\tau\nu_\tau$:} Hints for LFUV in these modes were observed first by the BaBar collaboration~\cite{Lees:2012xj} in 2012. These measurements have been confirmed by BELLE~\cite{Huschle:2015rga,Abdesselam:2016cgx} and LHCb has remeasured $B\to D^{*}\tau\nu_\tau$~\cite{Aaij:2015yra}. For the ratio ${R}(X)\equiv {\rm Br}[B\to X \tau \nu_\tau]/{\rm Br}[B\to X \ell \nu_\ell]$, the current HFAG average~\cite{Amhis:2014hma} of these measurements is
$
R(D)_\text{exp}=\,0.397\pm0.040\pm0.028  \,,\;\; 
R(D^*)_\text{exp}=\,0.316\pm0.016\pm0.010 \,.
$
Comparing these results to the SM predictions~\cite{Fajfer:2012vx} $R_\text{SM}(D)=0.297\pm0.017$ and 
$R_\text{SM}(D^*)=0.252\pm0.003$, there is a combined discrepancy of $4.0\sigma$~\cite{Amhis:2014hma}. 

{\boldmath $h\to\mu\tau$:} In the Higgs sector, CMS has presented results for a search for the lepton-flavor-violating (LFV) decay mode $h\to\mu\tau$, with a preferred value~\cite{Khachatryan:2015kon}
$	{\rm Br} [h\to\mu\tau] = \left( 0.84_{-0.37}^{+0.39} \right)\%.$
This is consistent with the less precise ATLAS measurement~\cite{Aad:2015gha}, giving a combined significance for NP of $2.6\sigma$, since such a decay is forbidden in the SM. This decay mode is of considerable interest because it hints at LFV in the charged-lepton sector, whereas up to now, LFV has only been observed in the neutrino sector via oscillations. 

{\boldmath $a_\mu$:} The AMM of the muon $a_\mu \equiv (g-2)_\mu/2$, provides another motivation for NP connected to muons. The experimental value of $a_\mu$ is completely dominated by the Brookhaven experiment E821~\cite{Bennett2006} and is given by $a_\mu^\mathrm{exp} = (116\,592\,091\pm54\pm33) \times 10^{-11}$, where the first error is statistical and the second systematic. The SM prediction is~\cite{Colangelo2014} $a_\mu^\mathrm{SM} = (116\,591\,855\pm59) \times 10^{-11}$, where almost the entire uncertainty is due to hadronic effects. This amounts to a discrepancy between the SM and experimental values of 
$
\Delta a_\mu = a_\mu^\mathrm{exp}-a_\mu^\mathrm{SM} = (236\pm 87)\times 10^{-11}\, ,
$
i.e.~a $2.7\sigma$ deviation\footnote{Less conservative estimates even lead to discrepancies up to $3.6\,\sigma$ in $a_\mu$}. 

\section{Explanations}

{\boldmath $b\to s\ell^+\ell^-$:} Here a flavour changing neutral current is required which can be naturally generated at tree-level by a $Z'$ vector bosons~\cite{Descotes-Genon:2013wba,Gauld:2013qba,Buras:2013dea,Altmannshofer:2014cfa,Crivellin:2015mga,Crivellin:2015lwa,Niehoff:2015bfa,Sierra:2015fma,Crivellin:2015era,Celis:2015ara} or by leptoquarks~\cite{Gripaios:2014tna,Becirevic:2015asa,Varzielas:2015iva,Alonso:2015sja,Calibbi:2015kma,Barbieri:2015yvd}. 

{\boldmath $B\to D^{(*)}\tau\nu_\tau$:} Here a tree-level NP contribution is required in order to generate the desired effect of the order of 25\% compared to the SM. Charged Higgses~\cite{Crivellin:2012ye,Tanaka:2012nw,Celis:2012dk,Crivellin:2013wna,Crivellin:2015hha} are one possibility, leading to large effects in the $q^2$ distribution. In addition, leptoquarks provide a valid explanation~\cite{Fajfer:2012jt,Deshpande:2012rr,Sakaki:2013bfa,Alonso:2015sja,Calibbi:2015kma,Bauer:2015knc,Fajfer:2015ycq,Barbieri:2015yvd} but also charged vector bosons are possible~\cite{Greljo:2015mma}. 

{\boldmath $a_\mu$:} NP in $b\to s\mu^+\mu^-$ should also contribute to the AMM of the muon. Explanations besides supersymmetry (see for example Ref.~\cite{Stockinger:2006zn} for a review) include leptoquarks~\cite{Chakraverty:2001yg,Cheung:2001ip}, new scalar contributions in two-Higgs-doublet models (2HDM)~\cite{Broggio:2014mna,Crivellin:2015hha}, and very light $Z^\prime$ bosons~\cite{Langacker:2008yv,Heeck:2011wj}

{\boldmath $h\to\mu\tau$:} Since the $B$ physics anomalies are related to $\tau$ and $\mu$ leptons, a connection to $h\to\mu\tau$ seems plausible. As the central value for the $h\to\mu\tau$ branching ratio is large, loop effects are in general not sufficient to generate the desired effect~\cite{Dorsner:2015mja}. Furthermore, also adding only vector-like fermions is not sufficient as the bounds from $\tau\to 3\mu$ and $\tau\to \mu\gamma$ are too stringent~\cite{Falkowski:2013jya}. Therefore, introducing additional scalars is the most popular option (see for example~\cite{Campos:2014zaa,Heeck:2014qea,Crivellin:2015mga,Dorsner:2015mja}). 

\begin{figure*}[t]
\centering
\includegraphics[width=0.7\textwidth]{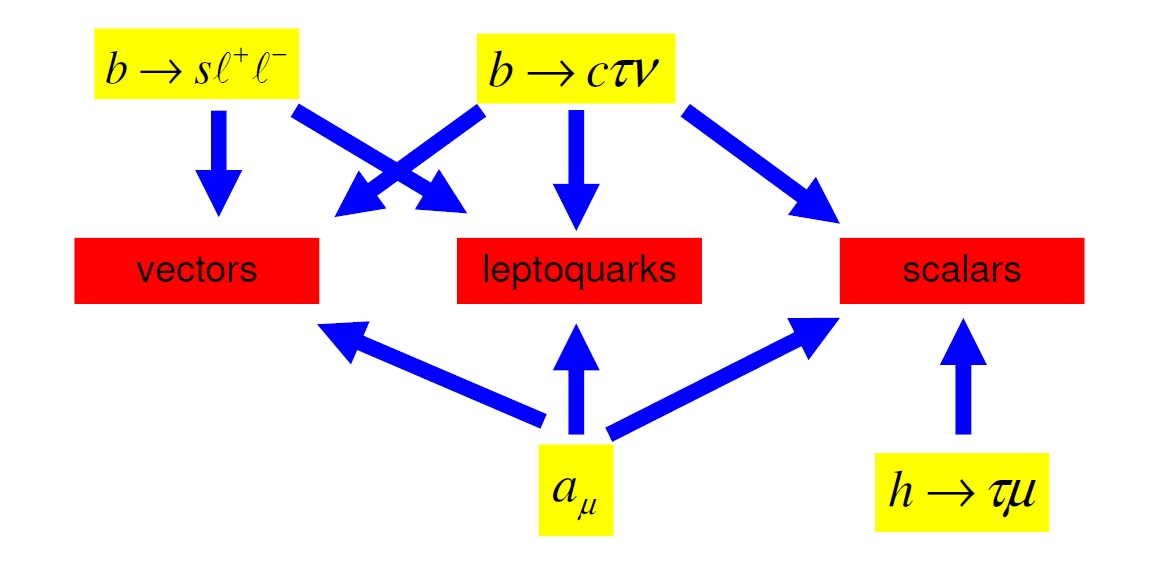}
\caption{Schematic picture of the implications for new particles from the various anomalies.\label{NPplot}}
\end{figure*}

\section{Selected models for simultaneous explanations of anomalies}

{\bf Multi Higgs {\boldmath  $L_\mu-L_\tau$} model: {\boldmath $h\to\tau\mu$} and {\boldmath $b\to s\mu^+\mu^-$} \cite{Crivellin:2015mga,Crivellin:2015lwa}}
\newline
Adding to a gauged $L_\mu-L_\tau$ model with vector like quarks~\cite{Altmannshofer:2014cfa} a second Higgs doublet with $L_\mu-L_\tau$ charge 2 can naturally give an effect in $h\to\tau\mu$ via a mixing among the neutral CP-even components of the scalar doublets. In this setup a $Z'$ boson, which can explain the $b\to s\mu^+\mu^-$ anomalies, gives sizable effects in $\tau\to3\mu$ which are potentially observable at LHCb and especially at BELLE II. One can avoid the introduction of vector-like quarks by assigning horizontal charges to quarks as well~\cite{Crivellin:2015lwa}. Then, the effects in $b\to s$, $b\to d$ and $s\to d$ transitions are related in an MFV-like way by CKM elements and the $Z^\prime$ can have an observable cross section at the LHC.
%\medskip

{\bf \boldmath Leptoquarks: $b\to s\mu^+\mu^-$ and $b\to c\tau\nu$~\cite{Calibbi:2015kma}}
\newline 
While in $b\to c\tau\nu$ both leptoquarks and the SM contribute at tree-level, in $b\to s\mu^+\mu^-$ one compares a potential tree-level NP contribution to a loop effect\footnote{Alternatively, there is one leptoquark representation for which one can explain $b\to s\mu^+\mu^-$ by a loop effect and $b\to c\tau\nu$ at tree-level in the case on anarchic couplings~\cite{Bauer:2015knc} and even explain the AMM of the muon.}. However, as $b\to c\tau\nu$ involves three times the third generation (assuming that the neutrino is of tau flavour in order to get interference with the SM contribution) but $b\to s\mu^+\mu^-$ only once. Therefore, leptoquarks with hierarchical flavour structure, i.e. predominantly coupling to the third generation~\cite{Glashow:2014iga,Bhattacharya:2014wla}, can explain simultaneously $b\to s\mu^+\mu^-$ and $b\to c\tau\nu$ in case of a $C_9=-C_{10}$ (left-handed quark and lepton current) solution for $b\to s\mu^+\mu^-$. In this case one predicts sizable effects in $B\to K^{(*)}\tau\tau$, $B_s\to\tau^+\tau^-$ and $B_s\to\mu^+\mu^-$ below the SM, while the effects in $b\to s\tau\mu$ are at most of the order of $10^{-5}$.
%\medskip

{\bf\boldmath 2HDM X: $a_\mu$ and $b\to c\tau\nu$ \cite{Crivellin:2015hha}}
\newline
In a 2HDM of type X, the couplings of the additional Higgses to charged leptons are enhanced by $\tan\beta$. As, unlike for the 2HDM II, this enhancement is not present for quarks, the direct LHC bounds on $H^0,A^0\to\tau^+\tau^-$ are not very stringent and also $b\to s\gamma$ poses quite weak constraints. Therefore, the additional Higgses can be light which, together with the $\tan\beta$ enhanced couplings to muons, allows for an explanation of $a_\mu$. If one adds couplings of the lepton-Higgs-doublet to third generation quarks, one can explain $b\to c\tau\nu$ as well by a charged Higgs exchange. In case of a simultaneous explanation of $a_\mu$ and $b\to c\tau\nu$ (without violating bounds from $\tau\to\mu\nu\nu$) within this model, sizable branching ratios (reaching even the \% level) for $t\to Hc$, with $m_H\approx 50-100\,$GeV and decaying mainly to $\tau\tau$, are predicted. Again, such a signature could be observed at the LHC.

%\medskip
{\bf\boldmath $L_\mu-L_\tau$ flavon model: $a_\mu$, $h\to\tau\mu$ and $b\to s\mu^+\mu^-$ \cite{Altmannshofer:2016oaq}}
\newline
In this model one adds vector-like leptons to the gauged $L_\mu-L_\tau$ model of Ref.~\cite{Altmannshofer:2014cfa}and one can explain $h\to\tau\mu$ via a mixing of the flavon (the scalar which breaks $L_\mu-L_\tau$) with the SM Higgs. Furthermore, one can account for $a_\mu$ by loops involving the flavon and vector-like leptons without violating the $\tau\to\mu\gamma$ bounds as this decay is protected by the $L_\mu-L_\tau$ symmetry. Despite the effects already present in the model of Ref.~\cite{Crivellin:2015mga}, one expects order one effects in $h\to\mu^+\mu^-$ detectable with the high luminosity LHC.

\section{Conclusions}
\label{conclusion}

In these proceedings we reviewed the anomalies in the flavour sector related to charged leptons together with some of their possible explanations. Interestingly, all anomalies involve muons and/or taus while the corresponding electron channels seem to agree with the SM predictions. This coherent picture of lepton flavour (universality) violation\footnote{For the implications in Kaon decays see~\cite{Crivellin:2016vjc}.} agrees with the stringent LEP constraints and suggests that the anomalies could be related, hinting at an unified explanation within a NP model. In Fig.~\ref{NPplot} we show in a schematic way which relations among the anomalies and new particles arise. Specific NP models can of course include the addition several new particles, potentially explain all anomalies and predict correlations among them and with other observables or processes detectable in future experiments.

\section*{Acknowledgments}
A.C. thanks the organizers for the invitation to \emph{Moriond QCD} and for the opportunity to present these results. This work is supported by an Ambizione grant of the Swiss National Science Foundation.
\bibliography{BIB}

\end{document}